\documentclass[aps,pre,reprint,twocolumn,superscriptaddress,showpacs]{revtex4-2}
\usepackage{amssymb,amsmath}
\usepackage[table]{xcolor}
\usepackage{graphicx}
\usepackage{mathtools}
\usepackage[export]{adjustbox}
\usepackage{overpic}
\usepackage[colorlinks=true,
linkcolor=black,
urlcolor=blue,
citecolor=blue]{hyperref}
\usepackage{nicefrac}
\usepackage{multirow}
\usepackage{lipsum} 
\usepackage[normalem]{ulem}
\usepackage{pbox}
\newcolumntype{C}[1]{>{\centering\let\newline\\\arraybackslash\hspace{0pt}}m{#1}}
\usepackage{verbatim}
\usepackage{enumitem}
\usepackage{caption}
\usepackage{subcaption}
\usepackage{float}

\usepackage{framed,color}

\allowdisplaybreaks
\begin{document}
\title{Mean-Field Analysis of the Glassy Dynamics of an Elastoplastic Model of Super-Cooled Liquids}
\author{Joseph W. Baron}
\email{joseph-william.baron@phys.ens.fr}
\affiliation{Laboratoire de Physique de l’Ecole Normale Sup\'{e}rieure, ENS, Universit\'{e} PSL, CNRS, Sorbonne Universit\'{e}, Universit\'{e} de Paris, F-75005 Paris, France}
\author{Giulio Biroli}
\email{giulio.biroli@phys.ens.fr}
\affiliation{Laboratoire de Physique de l’Ecole Normale Sup\'{e}rieure, ENS, Universit\'{e} PSL, CNRS, Sorbonne Universit\'{e}, Universit\'{e} de Paris, F-75005 Paris, France}

\begin{abstract}
We present a mean-field theory of a coarse-grained model of a super-cooled liquid in which relaxation occurs via local plastic rearrangements. Local relaxation can be induced by thermal fluctuations or by the long-range elastic consequences of other rearrangements. We extract the temperature dependence of both the relaxation time and the lengthscale of dynamical correlations. We find two dynamical regimes. First, a regime in which the characteristic time and length scales diverge as a power law at a critical temperature $T_c$. This regime is found by an approximation that neglects activated relaxation channels, which can be interpreted as akin to the one found by the mode-coupling transition of glasses. In reality, only a cross-over takes place at $T_c$. The residual plastic activity leads to a second regime characterised by an Arrhenius law below $T_c$. In this case, we show that the lengthscale governing dynamical correlations diverges as a power law as $T\to 0$, and is logarithmically related to the relaxation time. 
\end{abstract}

\maketitle

\section{Introduction}

Supercooled liquids experience a dramatic increase in viscosity (or, alternatively, relaxation time) as they approach the glass transition temperature. Along with this drastic slowing down, which can be by as much as 10 orders of magnitude for a 10\% drop in temperature \cite{angell1995formation}, the dynamics of the system becomes spatially heterogeneous \cite{berthier2011theoretical, berthier2011dynamical, berthier2005direct}. Furthermore, the spatial domains over which dynamics is correlated grow in size as the system is cooled, teasing the possibility of an underlying \textit{bona fide} phase transition. 
Exact solutions in large dimensions have demonstrated the possibility of a transition to a glassy ergodicty breaking state \cite{parisi2020theory,maimbourg2016solution}, which is consistent with a thermodynamic random first-order transition \cite{kirkpatrick1989scaling,biroli2021amorphous,biroli2022rfot}.

A rivalling view point contests that the primary mechanism responsible for the rapid increase of viscosity could be due to kinetic constraints \cite{chandler2010dynamics}. So called kinetically constrained models (KCMs) \cite{ritort2003glassy} have been employed to show how dynamic facilitation can give rise to dynamical heterogeneity. The phenomenon of dynamic facilitation, whereby relaxation in one region is made more likely by relaxation in a proximate region, is central in theoretical understandings of disordered media such as granular materials, colloids, and emulsions \cite{garrahan2011kinetically} (as was recently elucidated in \cite{scalliet2022thirty} using molecular dynamics simulations in a regime of deep super-cooling). Recent work on simulated molecular liquids has shown that effective local energy barriers grow as the system is cooled \cite{ciamarra2023energy}. A relaxation event, which occurs by deforming the surrounding solid, changes the energy barriers. This phenomenon offers a natural mechanism for the emergence of dynamical facilitation and dynamical heterogeneity \cite{chacko2021elastoplasticity,ozawa2023elasticity,tahaei2023scaling}, and an elasticity-based explanation of the presence of kinetic constraints \cite{hasyim2021theory}. A plausible complementary picture of super-cooled liquids is therefore as quasi-solids that flow due to thermally activated local rearrangements  \cite{dyre2006colloquium,lemaitre2014structural,dyre2023solid}.


Elasticity and plasticity play an important role in this local description of glassy dynamics. Elastoplastic models (EPMs) have been shown to be able to describe flow due to plastic rearrangements, where 
relaxation occurs due to Eshelby events \cite{lemaitre2014structural}, in absence of thermal fluctuations. In fact, EPMs have been shown to be effective descriptors of the yielding and flow of amorphous solids subject to external shear \cite{nicolas2017deformation}. The relevance of the elastoplastic picture to super-cooled liquids -- a case in which thermal fluctuations play a key role -- is bolstered by molecular simulations that have linked local yield stress with local rearrangements \cite{barbot2018local, zhang2021interplay}, and elastoplastic response with dynamical heterogeneity \cite{chacko2021elastoplasticity, lerbinger2022relevence}. Furthermore, dynamical heterogeneity has been seen to emerge naturally in thermal elastoplastic models \cite{ozawa2023elasticity}, with a spatial length-scale of dynamical correlations and thermal avalanches that grow with decreasing temperature \cite{tahaei2023scaling}.

In the following, we focus on thermal EPMs as models of super-cooled liquid dynamics \cite{ozawa2023elasticity}. 
The aim of our work is twofold:

On the one hand, we develop a mean-field analysis of dynamical heterogeneity in thermal EPMs. The main point of interest is the growth of the associated spatial length and how it is connected to the growth of the relaxation timescale. We follow an approach similar to that of Hebraud and Lequeux \cite{hebraud1998mode}, and Bocquet, Colin,  Ajdari \cite{bocquet2009kinetic} and generalize it to the thermal case. This allows us to study analytically the emergence of a growing lengthscale of dynamic heterogeneities with decreasing temperature. 

On the other hand, we show that the analysis based on thermal EPMs can be connected to other descriptions developed to analyze glassy dynamics. In fact, our mean-field analysis predicts two regimes of slow dynamics: a first one which is very reminiscent of the Mode-Coupling Theory (MCT) transition \cite{gotze1992relaxation}. Here the length increases as a power of relaxation time, and relaxation is not activated. Decreasing further the temperature, a cross-over takes place and dynamics becomes activated. In this second regime, studied in detail in \cite{tahaei2023scaling} in 2d thermal EPMs, and for which we provide a mean-field analysis, the length increases logarithmically with relaxation time. This regime is reminiscent of the one found and studied for Kinetically Constrained Models of glasses \cite{chandler2010dynamics}. 

Our analytical findings are verified by comparison to numerical simulations of the mean-field dynamics, with which we find excellent agreement.

\section{Elastoplastic model with thermally induced relaxation}
We consider a coarse-grained model of a supercooled fluid where space is discretised to form a hypercubic lattice, and each lattice point is assigned a local scalar stress $\sigma$. A full tensorial treatment of stress propagation is possible, but as was found in recent numerical treatments in finite dimensions \cite{tahaei2023scaling}, the same scaling laws and exponents were found in both the scalar and tensorial cases. 

We model each site as having a local energy barrier $\Delta E = K(\sigma_c -\sigma)^a$ (where $a \approx 1.5$) \cite{ferrero2019criticality,ferrero2021yielding, maloney2006energy, fan2014thermally, rodriguez2023temperature} that it must overcome in order to relax. When a site does relax, we say that a plastic rearrangement event has occurred. In the event of a rearrangement, the stress of the focal site is distributed to its neighbours via the Eshelby propagator \cite{eshelby1957determination, picard2004elastic}. There are two ways that a site may be provoked to relax: (i) thermal fluctuations may cause the site to overcome its local energy barrier, (ii) the stress `kicks' from the relaxation events at surrounding sites may increase the local stress above $\sigma_c$. In both cases, sites relax with a rate $1/\tau$. 

When sites relax, we assume that their stresses are redrawn from the following distribution \cite{popovic2020thermally, ozawa2023elasticity}
\begin{align}
	y(\sigma) = \frac{1}{\mathcal{N}}e^{(\vert \sigma\vert - \sigma_c)/\sigma_0}\theta(\sigma_c - \vert\sigma\vert) 
\end{align}
where $\theta(\cdot)$ is the Heaviside theta function and $\mathcal{N} = 2\sigma_0(1 - e^{-\sigma_c/\sigma_0})$. 

This can all be summarised by the following master equation for the probability $P_i(\sigma,t)$ of site $i$ having stress $\sigma$ at time $t$ \cite{bocquet2009kinetic}
\begin{align}
	\frac{\partial P_i}{\partial t} = \mathcal{L}(P, P) - \frac{1}{\tau} \nu(\sigma, \sigma_c) P_i + \Gamma_i(t) y(\sigma),\label{masterequation}
\end{align}
where $\mathcal{L}(P, P)$ encapsulates the inter-site interactions. The total rate of plastic events for sites of stress $\sigma$ is given by 
\begin{align}
	\nu(\sigma, \sigma_c) = \theta(\vert \sigma \vert - \sigma_c) + e^{-\frac{\Delta E(\sigma)}{T}}\theta(\sigma_c - \vert \sigma\vert), \label{ratenu}
\end{align}
and the total rate of plastic events (or `activity') $\Gamma_i(t)$ is determined self-consistently by
\begin{align}
	\Gamma_i(t) = \int_{-\infty}^{\infty} d\sigma \nu(\sigma, \sigma_c) P_i(\sigma, t). 
\end{align}
Specifically, $\mathcal{L}(P, P)$ encodes the transmission of stress between sites upon relaxation via randomly orientated Eshelby kernels \cite{ozawa2023elasticity}. A full expression for $\mathcal{L}(P,P)$ and a more detailed discussion is given in Appendix \ref{appendix:smallslopeapprox}. For the rest of this work, we will take $\sigma_c = \sigma_0 = K = \tau = 1$. 

\section{Mean-field theory and spatial heterogeneities}\label{section:mfandhet}
\begin{figure*}[t]
	\centering 
	\includegraphics[scale = 0.5]{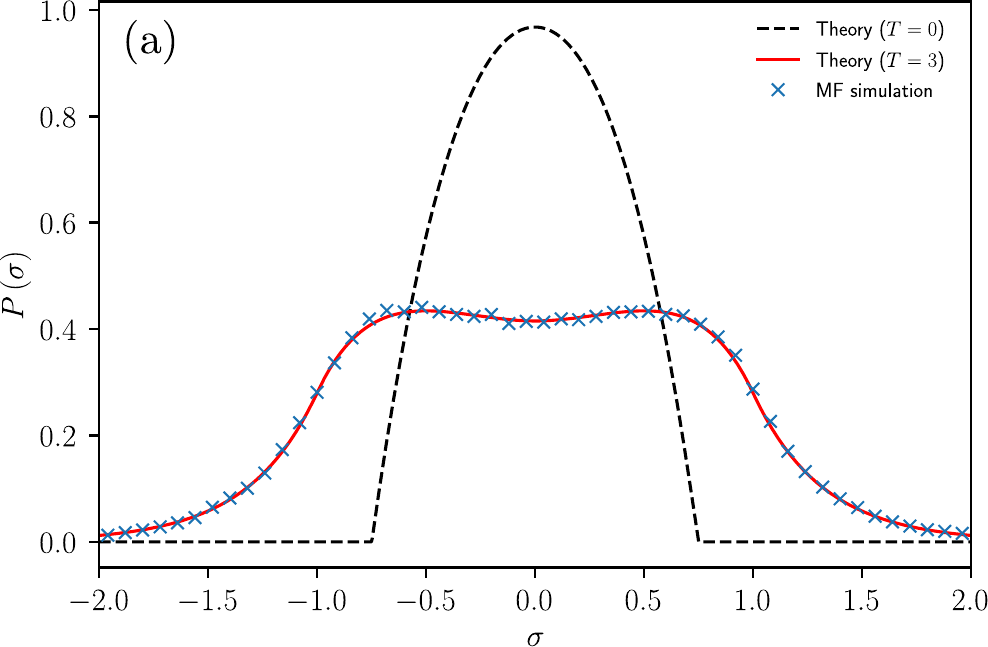}
	\includegraphics[scale = 0.5]{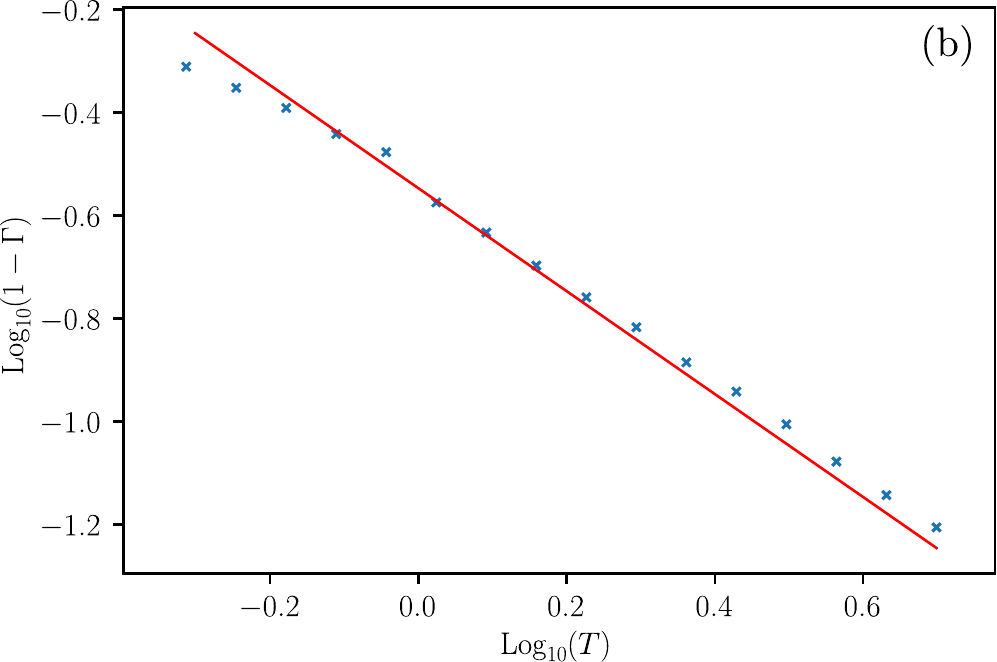}
	\captionsetup{justification=raggedright,singlelinecheck=false, font=small}
	\caption{Panel (a): Stationary distribution of stresses in the mean-field approximation. Blue crosses are the results of simulations as described in Appendix \ref{appendix:numerics} for $T = 3$ and $\alpha = 0.11$. The red line is the theory prediction found by solving Eq.~(\ref{fpeus}) using the approximation in Eq.~(\ref{nuhightapprox}). The dashed line is the result of Ref. \cite{ozawa2023elasticity} [also in Eq.~(\ref{tequal0})] for $T=0$, given for reference. Panel (b): Verification of the high temperature behaviour of the plastic activity $\Gamma(T)$. The solid line is the prediction in Eq.~(\ref{gammahight}). The blue crosses are the results of mean-field simulations. }\label{fig:statdisthight}
\end{figure*}

Following Refs. \cite{hebraud1998mode, popovic2020thermally, ozawa2023elasticity}, we make a mean-field assumption, which allows us to write a single, decoupled Fokker-Planck equation for the probability distribution $P_i(\sigma, t)$ for the stress at any one site at time $t$. 

First, we perform a Kramers-Moyal \cite{gardiner2009stochastic} expansion of $\mathcal{L}(P,P)$ [see Eq.~(\ref{lpp})], assuming that individual stress kicks are typically small, to obtain 
\begin{align}
	\frac{\partial P_i}{\partial t} = D_i(t)\frac{\partial^2 P_i}{\partial \sigma^2} - \frac{1}{\tau} \nu(\sigma, \sigma_c) P_i + \Gamma_i(t) y(\sigma), \label{fpegeneral}
\end{align}
By employing a small-slope argument (or derivative expansion)\cite{bocquet2009kinetic}, which should be a good approximation in this case where we expect the spatial heterogeneities to be slowly varying in space, we find that the `diffusion coefficient' satisfies (see Appendix \ref{appendix:smallslopeapprox})
\begin{align}
	D(\mathbf{r},t) = m \nabla^2 \Gamma(\mathbf{r},t) + \alpha \Gamma(\mathbf{r}, t), \label{spatialdependence}
\end{align}
where $\nabla^2$ denotes the Laplacian operator and we use a continuous spatial coordinate $\mathbf{r}$. Expressions for the coefficients $m$ and $\alpha$ in terms of the Eshelby kernel are given in Appendix \ref{appendix:smallslopeapprox}.
The diffusion coefficient encapsulates the noise induced by the rearrangements taking place in the surroundings of a particular site. 

One notes that in principle Eq.~(\ref{fpegeneral}) can be solved to yield $P(\sigma; \mathbf{r}, t)$ for each point in space, for arbitrary $D(\mathbf{r}, t)$ and $\Gamma(\mathbf{r}, t)$. Imposing a normalisation condition of $\int d\sigma P(\sigma;\mathbf{r}, t) = 1$ yields a constraint $Q(\Gamma(\mathbf{r},t), D(\mathbf{r},t)) = 0$. This constraint is the same for all points in space. The activity and the diffusion coefficient are only uniquely determined once we impose this `on-shell' condition in Eq.~(\ref{spatialdependence}).

Using this construction, which is a kind of Landau theory for EPMs, we now extract the mean-field behaviour plastic activity $\Gamma$ and the characteristic length-scale of its variation in space in various temperature regimes. In the following we find three dynamical regimes: a high-temperature one in which dynamics is not glassy, an MCT-like one in which dynamics is glassy but not activated, and finally an activated one. We will analyse each one of them separately. 

\section{High-Temperature behaviour}\label{section:hight}
We begin by briefly studying the high-temperature behaviour of the system. Understanding the behaviour for high temperatures will inform the approximations that we make to understand the MCT-like transition. For the time being, we leave aside any spatial dependence in $D$ and $\Gamma$, but we will reinstate it in the subsequent sections.

As one can see from Eq.~(\ref{ratenu}), the function $\nu(\sigma, \sigma_c)$ becomes increasingly flat as $T\to \infty$. We therefore adopt a linear approximation about $\vert\sigma\vert = 1/2$ for $\vert \sigma\vert <1$ to arrive at 
\begin{align}
	\nu(\sigma, \sigma_c) \approx e^{-\frac{1}{2^aT}}\left[1+ \frac{a}{2 ^{a-1}T} \left(\vert \sigma \vert -1/2\right) \right]. \label{nuhightapprox}
\end{align}
We demonstrate the accuracy of this approximation in Appendix \ref{appendix:highT}. 

Let us now obtain an approximate solution for the stationary profile $P(\sigma)$ and the corresponding rate of plastic events $\Gamma$ in the long-time limit. The uniform stationary solution $P(\sigma)$ satisfies
\begin{align}
	0&= D\frac{\partial^2 P}{\partial \sigma^2} - \frac{1}{\tau} \nu(\sigma, \sigma_c) P + \Gamma y(\sigma) ,\nonumber \\
	\Gamma &= \int d\sigma \nu(\sigma,\sigma_c) P(\sigma), \label{fpeus}
\end{align}
where we make the mean-field approximation $D = \alpha \Gamma$ [c.f. Eq.~(\ref{spatialdependence})]. 

Since we currently approximate $\nu(\sigma, \sigma_c)$ as the sum of two terms, the second of which is a small correction to the first, the solution for the stationary stress profile can be approximated by $P(\sigma) = P_0(\sigma) + P_1(\sigma)$, where $P_1/P_0 \sim a/(2^{a-1}T)$. 

The leading contribution $P_0(\sigma)$ is found by approximating $\nu \approx e^{-\frac{1}{2^aT}}$, i.e. constant in $\sigma$, and is relatively straight-forward to obtain. The correction $P_1$ can then be found in terms of $P_0$. This approximate solution, which contains only elementary functions, is given explicitly in Appendix \ref{appendix:highT}. As discussed in Section \ref{section:mfandhet}, one then imposes normalisation on this solution along with $D = \alpha \Gamma$ to find a spatially homogeneous value for $\Gamma$. An instance of the stationary solution is shown in Fig. \ref{fig:statdisthight}, where it is shown to agree well with numerical simulations of the mean-field theory.

From this mean field solution, one finds the following asymptotic behaviour for large $T$
\begin{align}
	\Gamma \sim 1 - \frac{T_{onset}}{T}, \label{gammahight}
\end{align}
where the temperature scale $T_{onset}$ delimits the high-temperature regime. We call it `onset' following Ref. \cite{sastry2000onset} that first introduced it for glass-forming liquids. For $T\gg T_{onset}$ dynamics is fast and the approximation above holds, whereas when the temperature decreases below $T_{onset}$ dynamic starts to slow down considerably and a new regime, studied in the section below, sets in. In Appendix \ref{appendix:highT} we give the explicit dependence on $\alpha$ of $T_{onset}$. For the value of $\alpha$ considered for thermal EPMs in Ref. \cite{ozawa2023elasticity} one finds $T_{onset}\simeq 0.284$. The high $T$ behaviour of $\Gamma(T)$ is verified in Fig. \ref{fig:statdisthight}b. 

\section{MCT-like Regime}\label{section:mct}
\begin{figure*}[t]
	\centering 
	\includegraphics[scale = 0.5]{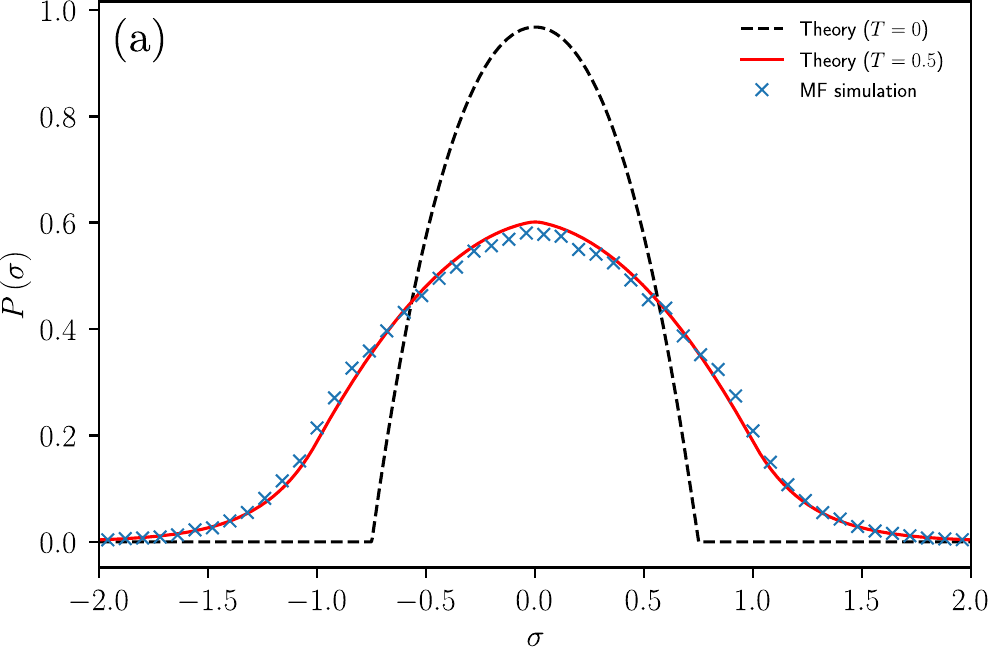}
	\includegraphics[scale = 0.5]{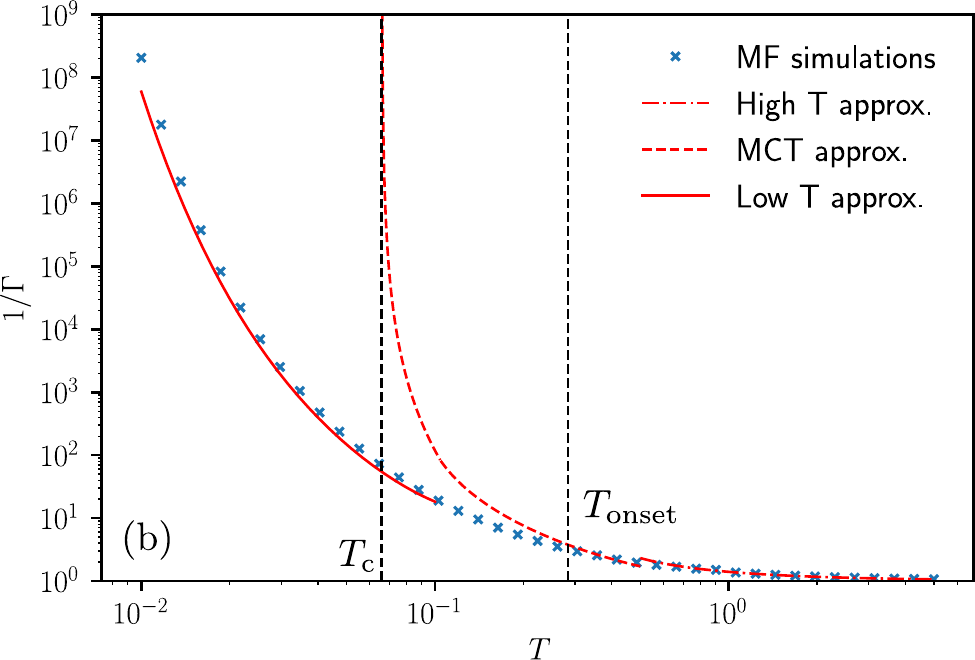}
	\captionsetup{justification=raggedright,singlelinecheck=false, font=small}
	\caption{Panel (a): Mean field stationary distribution of stresses. Blue crosses are the results of simulations as described in Appendix \ref{appendix:numerics} for $T = 0.5$ and $\alpha = 0.11$. The red line is the theory prediction found by solving Eq.~(\ref{fpeus}) using the approximation in Eq.~(\ref{barrierapprox}). The dashed line is again the result of Ref. \cite{ozawa2023elasticity} [also in Eq.~(\ref{tequal0})] for $T=0$. Panel (b): Theoretical predictions for the relaxation time $1/\Gamma$ in the different regimes. We see that while the approximation in Eq.~(\ref{barrierapprox}) is reasonably effective for $T\approx 0.1$ to $T\approx0.5$, it breaks down around $T_c\approx 0.07$.  }\label{fig:statdistmc}
\end{figure*}

We now consider the first regime of dynamical slowing down. Simulations of super-cooled liquids have shown that in this first regime dynamics is mainly not activated \cite{berthier2011theoretical}. Yet, by decreasing the temperature, the relaxation times increase by several orders of magnitude. This regime, related to the formation of metastable states, has been described with some success by the Mode-Coupling Theory of glasses (MCT) \cite{reichman2005mode, leutheusser1984dynamical, bengtzelius1984dynamics}. 

Our analysis reveals the existence of a similar regime in thermal EPMs. In order to characterise it, we approximate the function $\nu(\sigma, \sigma_c)$ as if {\it only} relaxations that are not strongly thermally activated are allowed to occur. That is, for relaxations that are strongly thermally activated, i.e. for which $\nu(\sigma, \sigma_c)$ is exponentially small in $1/T$, we set the rate to zero. 

In practice, we do this by linearising about the point where $\nu(\sigma, \sigma_c) = 1/2$. We say that values of $\sigma$ for which $\nu(\sigma, \sigma_c)<0$ in this approximation are `strongly activated', and we set $\nu(\sigma, \sigma_c) =0$ for these values. In summary, our approximation is
(for $\vert\sigma\vert<1$)
\begin{align}
	\nu(\sigma, \sigma_c) &\approx \Theta(\vert\sigma\vert - \sigma_B(T))\left[1/2 + \Delta (\vert\sigma\vert - \sigma_{1/2}) \right], \nonumber \\
	\Delta &= \frac{a}{2 T} \left[\ln(2) T \right]^{\frac{a-1}{a}}, \nonumber \\
	\sigma_{1/2} &= 1- \left[\ln(2) T \right]^{\frac{1}{a}}, \label{barrierapprox}
\end{align}
where $\sigma_B(T) = \sigma_{1/2}- 1/(2 \Delta)$ is the value of $\vert\sigma\vert$ at which the linear approximation hits zero, and $\Theta(\cdot)$ is the Heaviside function. 

In the following, we shall analyze the dynamical behavior resulting from cutting out strongly activated relaxation. 
As we will see, the stationary distribution of stresses $P(\sigma)$ (in the mean-field approximation) will extend into the region $\sigma_B < \sigma < 1$ when the temperature satisfies $T>T_c \approx 0.07$. However, as $T$ is reduced, the local energy barriers increase, hence only few of them are still of order $T$ and do not require strongly activated thermal relaxation. The rarefaction of these non-activated relaxation channels induces a slowing down of the dynamics, and eventually an MCT-like transition \cite{hebraud1998mode}. 

According to the approximation in Eq.~(\ref{barrierapprox}), the stationary distribution will adopt a frozen state \cite{hebraud1998mode} where $P(\sigma) >0$ only for $\vert \sigma \vert <\bar\sigma$ with $\bar\sigma<\sigma_B$ so that no thermal excitations can occur. As we shall show, this transition is accompanied by a growing length-scale with the approach to $T_c$. The growth of time and length are both power laws, as for MCT and similarly to what found in simulations of super-cooled liquids \cite{berthier2011dynamical}.  
At the transition point, in reality there is a cross-over where the approximation we made in Eq.~(\ref{barrierapprox}) is no longer reasonable. This is exactly what is also found for MCT. At lower temperature, one instead enters a new regime where the rearrangements that matter for dynamics are thermally activated.

\subsection{Mean-field stationary distribution}
Proceeding as in Section \ref{section:hight}, we can solve Eq.~(\ref{fpegeneral}) in a piece-wise fashion for the regions $0<\sigma<\sigma_B$, $\sigma_B<\sigma<1$ and $\sigma>1$. Any constants of integration are once again determined by imposing that the function $P(\sigma)$ is continuous and smooth at the points $\vert\sigma\vert = \sigma_B$ and $\vert\sigma\vert = 1$, and also that the derivative at $\sigma = 0$ is nil (to ensure the correct symmetry). 

We use two different approximation schemes for $P(\sigma)$, in conjunction with that in Eq.~(\ref{barrierapprox}), depending on the temperature. In order to understand the approach to the mode coupling temperature $T_c$, we make a similar expansion to the previous section, but taking $\bar\sigma - \sigma_B$ as the small parameter. For larger values of $T$, we instead find a series solution in $\sigma$ for $P(\sigma)$ about $\sigma_{1/2}$. 

An example of the mean-field stationary distribution of stresses $P(\sigma)$, using the latter approximation scheme, is given in Fig. \ref{fig:statdistmc}. We use the former approximation to give a more precise analytical account of the approach to $T_c$. We give a more detailed discussion and explicit expressions for $P(\sigma)$ in Appendix \ref{appendix:mct}.

\subsection{Scaling of relaxation time for $T \rightarrow T_c$}

\begin{figure*}[t]
	\centering 
	\includegraphics[scale = 0.5]{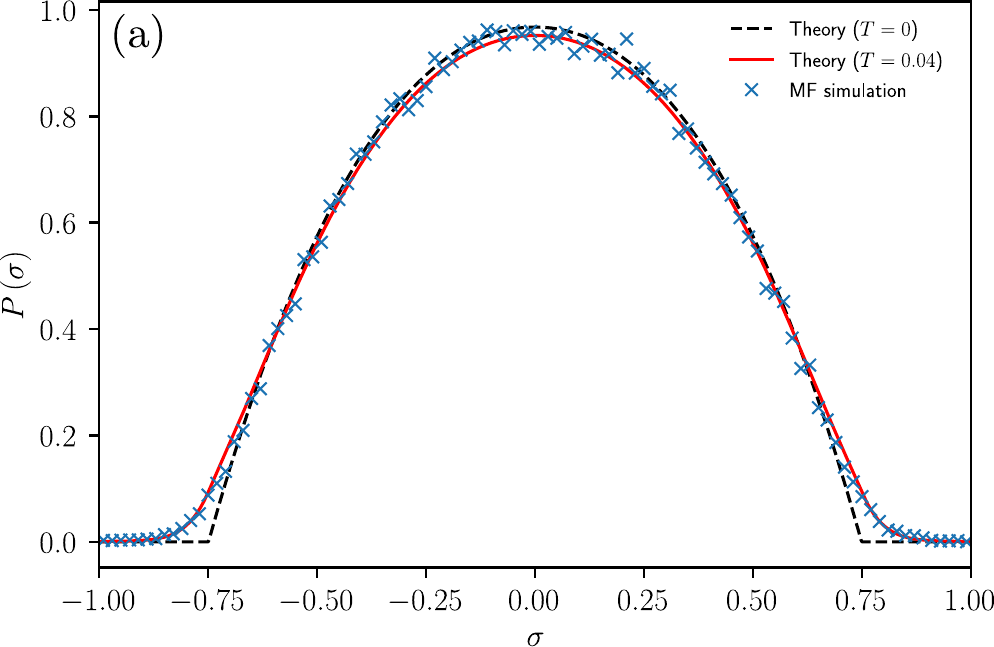}
	\includegraphics[scale = 0.5]{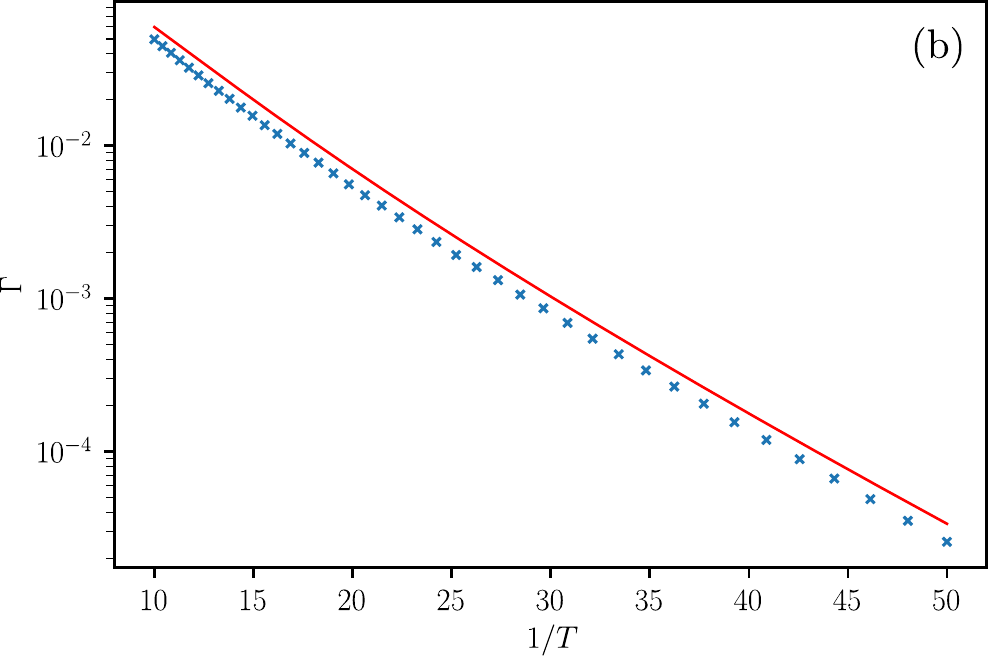}
	\captionsetup{justification=raggedright,singlelinecheck=false, font=small}
	\caption{Panel (a): The stationary distribution $P(\sigma,T)$. The solid red curve is the approximate low-$T$ solution given in Eqs.~(\ref{insidesolution}) and (\ref{outsidesolution}). The dashed black curve curve is the $T \to 0$ solution in Eq.~(\ref{tequal0}). The crosses are the results of simulating the mean-field dynamics using the procedure described in Appendix \ref{appendix:numerics}, with $N = 1000$, averaged over 10 trials. The temperature is $T = 0.04$ and $\alpha = 0.11$.  Panel (b): Scaling of the rate of plastic events $\Gamma$ with temperature as $T\to 0$ [see Eq.~(\ref{gammascaling})]. Mean field simulation results are averaged over 10 trials. }\label{fig:statdist}
\end{figure*}

We once again proceed in the same way as Section \ref{section:meanfieldT0} to deduce $\Gamma(T)$ (the inverse relaxation time) and $D(T)$ in the mean field approximation. 

For $T \approx T_c$, the rate of plastic deformations $\Gamma$ (and consequently $D$) is small. Using this fact, we can approximate the condition $Q(\Gamma, D) = 0$ (see Section \ref{section:mfandhet}) as
\begin{align}
	D \approx& \alpha \Gamma + \beta_1 [\sigma_B(T)- \bar \sigma] \Gamma + \frac{\beta_2(T) }{[\sigma_B(T) - \bar\sigma]^2} D \Gamma, \label{gammaexpansion}
\end{align} 
where the expressions for $\beta_1$ and $\beta_2$ are given in Appendix \ref{appendix:mct}. Now, setting $D = \alpha \Gamma$, we find two solutions for $\Gamma$
\begin{align}
	\Gamma = 
	\begin{cases}
		-\frac{\beta_1}{\beta_2(T)}[\sigma_B(T) - \bar \sigma]^3 & \text{if } T>T_c\\
		0 & \text{if } T\leq T_c .
	\end{cases}\label{homsol}
\end{align}

The critical temperature is seen to be defined by $\sigma_B(T_c) = \bar\sigma$. We can now deduce the scaling of $\Gamma$ on the approach to the critical point. One finds that the temperature dependence of $\beta_2(T)$ is slow compared with $\sigma_B(T)- \bar\sigma$. Hence, we have 
\begin{align}
	\Gamma \sim [\sigma_B(T) - \sigma_B(T_c)]^3. \label{gammascalingmctsigma}
\end{align}
The function $\sigma_B(T)$ has a non-zero derivative at $T_c$, and hence, one finally arrives at 
\begin{align}
	\Gamma \sim [T-T_c]^3.
\end{align}
That is, our analysis predicts a power law divergence of the relaxation times with an exponent three. This MCT-like prediction for $\Gamma(T)$ is compared to simulation results for the mean field model in Fig. \ref{fig:statdistmc}b. 
It is reminiscent of similar plots for super-cooled liquids \cite{berthier2011theoretical}.
In fact, when comparing to experimental data, the MCT transition temperature is inferred as the temperature where a power law divergence in the viscosity would occur according to a fit for higher temperatures. In reality, a divergence does not occur at this temperature. Instead one finds a residual finite (but large) viscosity.
Similarly, the approximation in Eq.~(\ref{barrierapprox}) reflects the first regime of slowing down of the dynamics. It predicts a power-law divergence at $T_c \approx 0.07$, which is actually just a cross-over. 

\subsection{Growing length-scale}\label{section:lengthscalemct}

Following Ref. \cite{bocquet2009kinetic}, we can also use Eq.~(\ref{gammaexpansion}), which applies at every point in space, to identify a growing length scale as $T \to T_c$. We write $\Gamma(\mathbf{r}) = \Gamma^\star + \delta\Gamma(\mathbf{r})$ [and similarly for $D(\mathbf{r})$], and we use Eq.~(\ref{spatialdependence}), where $\Gamma^\star$ and $D^\star$ are the mean-field solutions that simultaneously satisfy Eq.~(\ref{gammaexpansion}) and $D^\star = \alpha\Gamma^\star$. One thus finds for small $\delta\Gamma(\mathbf{r})$ and $\delta D(\mathbf{r})$
\begin{align}
	\nabla^2 \delta \Gamma = -\frac{1}{\xi^2}\delta \Gamma,
\end{align}
where we have identified the lengthscale 
\begin{align}
	\xi = \sqrt{\frac{m}{\vert\beta_1 (\sigma_B(T) - \bar\sigma) \vert}} \sim (T-T_c)^{-1/2}.
\end{align}
This lengthscale governs the way in which perturbations affect dynamics. It is similar in spirit to the one obtained by inhomogeneous MCT \cite{biroli2006inhomogeneous}. 
In summary, we have found that as the critical temperature is approached, the relaxation time $1/\Gamma$ diverges and so does the length-scale associated to dynamical correlations, according to the linear approximation in Eq.~(\ref{barrierapprox}). The divergences are both power laws, as in MCT. The values of the exponents are different though. Within MCT, the one associated with the relaxation timescale is system dependent (although often close to three), whereas the one associated to the length is $1/4$ \cite{biroli2006inhomogeneous}. It would be interesting to investigate whether a more refined dynamical mean-field approximation would be able to reproduce these values. 

\section{Zero-temperature transition}\label{section:T0}

In the previous section, we studied the first regime of slowing down of the dynamics, which is not due to strongly activated events. Our analysis consisted of cutting off relaxation channels with rates exponentially small in $1/T$ and performing a linear approximation of the function $\nu(\sigma,\sigma_c)$ for those that remained. This approximation predicted an MCT-like transition at a temperature $T_c$. In reality, this is just a cross-over since there is instead a small amount of residual activity even for $T<T_c$. 

In this section, we analyze the regime below $T_c$ using a systematic expansion for small $T$. Our aim is to obtain a mean-field analysis of the dynamics for $T\to 0$. Also in this regime, we find that time and length-scales grow, but their relationship changes from the MCT regime; the time grows exponentially with $1/T$, whereas the lengthscale grows as a power law. As a consequence, the characteristic length-scale goes as a power of the logarithm of the relaxation time. Experiments have been seen to give evidence of such a cross-over \cite{berthier2005direct}.

\subsection{Mean-field stationary distribution for small $T$}\label{section:statdistT0}
We first present the $T\to 0$ uniform stationary solution to Eqs.~(\ref{fpegeneral}) derived in Ref. \cite{ozawa2023elasticity}. In this work, the coefficient $\alpha$, which determines whether the system is in the `frozen' state \cite{hebraud1998mode}, was taken to be $\alpha = 0.110$, so as to agree with the Eshelby kernel evaluated numerically on the 2-d square lattice. This meant that the system was below the 'frozen' threshold, and the zero-temperature stationary state would be $\Gamma = 0$. In principle, the stationary distribution of stresses in the frozen state depends on the initial conditions \cite{hebraud1998mode, parley2020aging}. However, in the limit of small but finite $T$ there is always some small degree of thermal activity, so that eventually the system always tends towards a symmetric distribution of stresses in the long run. 

The uniform stationary solution $P(\sigma)$ again satisfies Eq.~(\ref{fpeus}). Given the symmetry constraint $P(\sigma) = P(-\sigma)$, the stationary distribution $P(\sigma)$ approaches in the limit $T\to 0$ \cite{ozawa2023elasticity}
\begin{align}
	P_0(\sigma) =& \bigg[\frac{\sigma_0^2}{\alpha \mathcal{N}}\left(e^{(\bar\sigma - \sigma_c)/\sigma_0} - e^{(\vert\sigma\vert - \sigma_c)/\sigma_0}\right) \nonumber \\
	&+ \frac{\sigma_0 e^{-\sigma_c/\sigma_0}}{\alpha \mathcal{N} }(\vert\sigma\vert-\bar \sigma) \bigg]\theta(\bar\sigma - \vert \sigma \vert).\label{tequal0}
\end{align}
That is, the stationary probability distribution has a finite support in the limit $T\to 0$. Imposing normalisation, one sees that the value of boundary of the support $\bar \sigma$ satisfies 
\begin{align}
	\alpha = \frac{\sigma_0(\bar \sigma - \sigma_0)e^{(\bar\sigma-\sigma_c)/\sigma_0}+ (\sigma_0^2 -\bar\sigma^2/2)e^{-\sigma_c/\sigma_0}}{1 - e^{-\sigma_c/\sigma_0}} .
\end{align}
Solving numerically, one finds $\bar\sigma \approx 0.749$. We now seek the lowest-order correction to the stationary distribution for small $T$. This is accomplished by assuming that most thermal activations and plastic events occur in the vicinity of $\sigma \approx \bar\sigma$ (in a similar fashion to Ref. \cite{popovic2020thermally}). This allows us to approximate the energy barrier with a linear expression to obtain the following approximation for $0<\sigma<\sigma_c$
\begin{align}
	\nu(\sigma, \sigma_c) \approx e^{-\bar E /T} e^{-\frac{(\bar\sigma - \sigma)}{T_1}}, \label{nusmallt}
\end{align}
where $\bar E = (\sigma_c - \vert \bar \sigma \vert)^a$ and $T_1 = T/[a (\sigma_c - \bar \sigma)^{a-1}]$. 

We now identify two regimes for $\sigma$. These are: 1) $0<\sigma<\bar\sigma$, in which the Arrhenius term in Eq.~(\ref{nusmallt}) is small in comparison to $\Gamma$, and 2) $\sigma>\bar\sigma$, in which the Arrhenius term is large compared to $\Gamma$. One can neglect the contributions to $P(\sigma)$ from the range $\sigma>\sigma_c$. 

As is shown in Appendix \ref{appendix:smallt}, by once again solving Eq.~(\ref{fpeus}) in these two regions separately and requiring continuity and smoothness of $P(\sigma)$ at $\sigma = \bar \sigma$ and normalisation, we obtain a piecewise solution for $P(\sigma)$. We can thus deduce $\Gamma(T)$ using the mean-field assumption $D = \alpha \Gamma$ as described in the previous sections. An example of the stationary solution, using the approximation in Eq.~(\ref{nusmallt}), is shown in Fig. \ref{fig:statdist}.

\subsection{Scaling of relaxation time for low $T$}\label{section:meanfieldT0}

In a similar way to Section \ref{section:mct}, we now approximate the constraint $Q(\Gamma, D) = 0$ close to the transition point at $T = 0$. As we discuss in more detail in Appendix \ref{appendix:smallt}, one observes that, upon defining $d = D e^{\bar E/T}$ and $\gamma = \Gamma e^{\bar E/T}$, we can rewrite the constraint as $q(\gamma, d) = 0$, where $q(\gamma, d)$ is now a rational expression in $T$. We can expand this condition for small $T$ (which connotes small $d$ and $\gamma$) to find
\begin{align}
	\gamma \approx \left( \frac{1}{\alpha} + c_{1}T_1 \right) d +  c_{2}T_1  d^{3/2} +  c_{3}  d^2 ,\label{constraintlowt}
\end{align}
where the constants $\{c_{a}\}$ are given in Appendix \ref{appendix:smallt}. By imposing $d = \alpha \gamma$, we arrive at 
\begin{align}
	\gamma \approx -\frac{1}{\alpha}\frac{c_{1}}{c_{3}}\frac{T}{a (1 - \bar \sigma)^{a-1}}   \label{prefactorscaling}
\end{align}
for $T\to 0$. So, we find for the scaling of $\Gamma$ in the mean-field approximation
\begin{align}
	\Gamma \sim T e^{-\bar E/T} .\label{gammascaling}
\end{align}
This implies an Arrhenius law with a prefactor $1/T$ for the relxation times. A  similar behaviour was also found numerically in Ref. \cite{rodriguez2023temperature}. The direct verification of the scaling of the prefactor is beyond the immediate scope of our numerical simulations, since the equilibration time increases drastically as the temperature is lowered. However, we compare the theoretical prediction for $\Gamma(T)$ against simulations to low temperatures that are accessible to us in simulations in Fig. \ref{fig:statdist}b, where we observe the dominant Arrhenius law behaviour. We subsequently verify the prefactor scaling in Eq.~(\ref{gammascaling}) by solving $q(\gamma,d)=0$ and $d = \alpha \gamma$ numerically for smaller values of $T$ (see Appendix \ref{appendix:smallt}). As shown in \cite{ozawa2023elasticity} this result is also a good approximation for the finite dimensional system.

\subsection{Diverging length-scale}
As we did in Section \ref{section:lengthscalemct}, we can also find the temperature dependence of the lengthscale of spatial inhomogeneities. Once again, we assume that inhomogeneities are slowly varying in space, and we solve Eq.~(\ref{constraintlowt}) (which is valid at each point in space) simultaneously with Eq.~(\ref{spatialdependence}). 

Specifically, we linearise about the homogeneous solution $\Gamma(T)$ and $D(T)$ that we found in Section \ref{section:meanfieldT0} and define $\delta \gamma = e^{\bar E/T}[\Gamma(\mathbf{r}) - \Gamma]$ and  $\delta d = e^{\bar E/T}[D(\mathbf{r}) - D]$.  Ultimately, we find for $T \to 0$ 
\begin{align}
	\Delta(\delta \gamma) &\approx  \frac{\alpha^2 c_1 T_1}{m}\delta \gamma,
\end{align}
from which we extract the length-scale temperature dependence
\begin{align}
	\xi \sim T^{-\frac{1}{2}} . \label{lengthscaleT0}
\end{align}
The divergence of the dynamical correlation length $\xi$ implies that thermal EPMs display a zero-temperature transition. The associated criticality was studied by numerical simulations and scaling theory in  \cite{tahaei2023scaling}.

\section{Discussion}
In summary, we have presented a mean-field Landau-like analysis of thermal elastoplastic models of super-cooled liquids. Our main result is showing the existence of two dynamical regimes: a first one in which energy barriers are still of the order of the temperature and dynamics slows down because of rarefaction of relaxation channels. This regime has many properties in common with the one associated to MCT; time and length-scales increase as an inverse power law of the distance to the transition, and the transition is actually just a cross-over toward strongly activated dynamics. 

For the second regime, already studied by simulations and scaling theory in \cite{ozawa2023elasticity,tahaei2023scaling}, our mean-field treatment predicts a diverging length-scale of dynamical correlation as $1/\sqrt{T}$, and hence a zero-temperature critical point. As found in simulations, in this regime length and time scales are logarithmically related. However, we note that, in Ref. \cite{tahaei2023scaling}, the temperature dependence of the characteristic length-scale associated to thermal avalanches and dynamical correlations was found to satisfy $\xi \sim T^{-\delta/d_f}$ for $T\to 0$, where $d_f$ is related to the spatial dimension of the system. Our result, presented in Eq.~(\ref{lengthscaleT0}), is instead manifestly independent of the spatial dimension. This is because our treatment in inherently mean-field like and can only capture qualitatively the critical behavior of two-dimensional systems.  

There are several directions that are worth further work. First, it would be interesting to consider a more refined mean-field analysis. In particular, 
one key aspect of our approach here was to assume that distribution of stress kicks $\rho(\delta\sigma,t)$ permitted a series expansion in small $\delta\sigma$. In Ref. \cite{parley2020aging}, a kernel that took into account the possibility of large stress kicks was used. We imagine that the exponent of $1/2$ in Eq.~(\ref{lengthscaleT0}) could be altered by such an accommodation, just as the exponent of the athermal relaxation of $\Gamma$ with time was affected in Ref. \cite{parley2020aging}. Moreover, there are different ways to perform the mean-field analysis. We followed the Kramers-Moyal expansion developed in \cite{bocquet2009kinetic}, but other more refined procedures can be envisaged, e.g. one based on DMFT as done for the analysis of high-dimensional super-cooled liquids \cite{maimbourg2016solution}. Finally, it would be very interesting to develop a framework to go beyond mean-field theory and obtain a first-principle theory of the finite dimensional criticality studied by scaling theory and simulations in \cite{tahaei2023scaling}. 

Overall, our results provide additional evidence that thermal EPMs offer a new and interesting framework to study the glass transition. These models put flesh on the description of super-cooled liquids as solids that flow \cite{dyre2006colloquium,lemaitre2014structural,dyre2023solid}, and at the same time they connect to previous descriptions such as MCT \cite{gotze1992relaxation} and dynamical facilitation by kinetic constraints \cite{chandler2010dynamics}.

\acknowledgements
This work was supported by grants from the Simons Foundation (\#454935 Giulio Biroli). We would like to acknowledge useful discussions with M. Ozawa.

\appendix

\begin{widetext}

\setcounter{figure}{0}		

\renewcommand{\thefigure}{S\arabic{figure}}  		

\newpage
\pagebreak
	
\section{Small-slope approximation for the space-dependence of the activity}\label{appendix:smallslopeapprox}
To arrive at Eq.~(\ref{fpegeneral}), we first begin with the expression for the stress propagator in Eq.~(\ref{masterequation})
\begin{align}
	\mathcal{L}(P,P) =&  \sum_{j\neq i} \int d(\delta\sigma_j) \Gamma_j \rho_j(\delta \sigma_j, t)\bigg[ P_i(\sigma + G_{ij}^{\psi_j} \delta \sigma_j, t) - P_i(\sigma,t)\bigg] , \label{lpp}
\end{align}
where $\rho_j(\delta\sigma_j, t)$ is the probability that, given that site $j$ experiences a stress drop at time $t$. The stress drop is of size $\delta\sigma_j$, and $G_{ij}^{\psi_j}$ is the Eshelby stress propagator with random orientation $\psi_j$. If the material that we were modelling were subject to a macroscopic shear in a particular direction, the orientations of the Eshelby kernels would be aligned with that direction \cite{maloney2006amorphous, dasgupta2013yield}. However, in our case we consider no shear. The medium is instead modelled as disordered. Interactions between different locations have no preferred direction and can change between relaxation events \cite{ozawa2023elasticity, lemaitre2014structural, wu2015anisotropic}. 

Performing a Kramers-Moyal \cite{gardiner2009stochastic} expansion of $\mathcal{L}(P,P)$ [see Eq.~(\ref{lpp})], assuming that individual stress kicks are typically small, one obtains the Fokker-Planck equation in Eq.~(\ref{fpegeneral}), where the `diffusion coefficient' is given by
\begin{align}
	D_i(t) =& \frac{1}{2}\sum_{j\neq i} \int d(\delta\sigma_j) \Gamma_j \, \rho_j(\delta \sigma_j, t) \, \left(G_{ij}^{\psi_j} \delta \sigma_j\right)^2 .
\end{align}
Now, we assume that the average squared stress drop is the same for all sites to obtain
\begin{align}
	D_i(t) \approx& \frac{1}{2} \sum_{j\neq i} \left(G_{ij}^{\psi_j}\right)^2 \Gamma_j(t) \left\langle \left( \delta \sigma\right)^2 \right\rangle. \label{di}
\end{align}
We now perform a second expansion to obtain the leading order spatial dependence of $D_i(t)$. Following Ref. \cite{bocquet2009kinetic}, we make a small-slope approximation (which is valid when $\Gamma_i$ varies on a much larger scale than the lattice spacing). If we write $\delta r^\alpha_{ij} = r^{\alpha}_j - r^\alpha_i$ for the $\alpha$ spatial component of the vector displacement between sites, we obtain the following upon expanding $\Gamma_j$ as a function of its spatial coordinate 
\begin{align}
	D_i( t) &\approx \frac{1}{2}  \left\langle \left( \delta \sigma\right)^2 \right\rangle \sum_{j\neq i} \left[G^{\psi_j}_{ij}\right]^2 \left[ \Gamma_i + \sum_{\alpha \beta} \delta r_{ij}^\alpha \delta r_{ij}^\beta \frac{\partial^2 \Gamma_i}{\partial (\delta r_{ij}^\alpha) \partial (\delta r_{ij}^\beta) }  \right], \nonumber \\
	&\approx \frac{1}{2}  \left\langle \left( \delta \sigma\right)^2 \right\rangle \sum_{j\in nn(i)} \left[G^{\psi_j}_{ij}\right]^2 \left( \Gamma_i +  a^2 \Delta \Gamma_i \right)
\end{align}
where by the rotational symmetry of the Eshelby kernel, various terms in the expansion of $\Gamma_i(\mathbf{r}_i + \delta \mathbf{r}_{ij})$ have vanished and we have approximated only the nearest neighbours on the square lattice as contributing to the sum. We use $a$ to denote the lattice spacing. One thus arrives at Eq.~(\ref{spatialdependence}) in the main text with
\begin{align}
	m &= \frac{a^2}{4} \left\langle \left( \delta \sigma\right)^2 \right\rangle  \sum_{j \in nn(i)} \left(G_{ij}^{\psi_j}\right)^2  , \nonumber \\
	\alpha &= \frac{1}{2} \left\langle \left( \delta \sigma\right)^2 \right\rangle  \sum_{j\neq i} \left(G_{ij}^{\psi_j}\right)^2 , \label{spatialcoefficients}
\end{align}
where $nn(i)$ denotes the set of nearest neighbours of site $i$. 

\section{Stationary profile for high $T$}\label{appendix:highT}
To find the approximation for the stationary distribution of local stresses $P(\sigma)$ in the mean-filed limit for large $T$, we begin with Eq.~(\ref{fpeus}) and use the mean-field approximation that $D = \alpha \Gamma$ is uniform in space [c.f. Eq.~(\ref{spatialdependence})].

As mentioned in Section \ref{section:hight}, we assume that the function $\nu(\sigma, \sigma_c)$ is approximately linear for $\vert \sigma \vert <1$. More precisely, we write $\nu \approx \nu_0 + \nu'_0(\vert\sigma\vert -1/2)$ where we have $\nu_0 = e^{-\frac{1}{2^a T}}$ and we assume that $\nu'_0= e^{-\frac{1}{2^a T}}\frac{a}{2^{a-1}T}$ is small. This approximation is shown in Fig. \ref{fig:nuapproxhight} below.
\begin{figure}[H]
	\centering 
	\includegraphics[scale = 0.5]{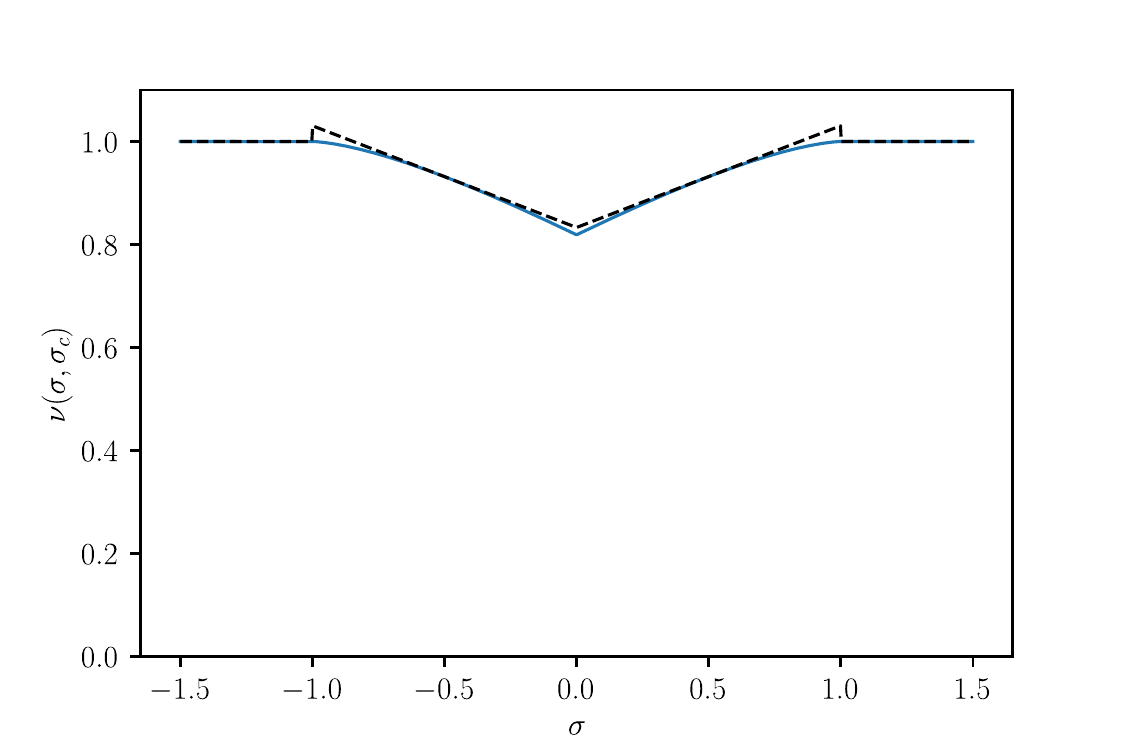}
	\captionsetup{justification=raggedright,singlelinecheck=false, font=small}
	\caption{Comparison of the approximate expression in Eq.~(\ref{nuhightapprox}) with Eq.~(\ref{ratenu}) for $T = 3$. }\label{fig:nuapproxhight}
\end{figure}
We now write a series approximation for $P(\sigma) \approx P_0(\sigma) + P_1(\sigma)$, where $P_0$ is the solution that we obtain by setting $\nu'_0 = 0$, and we assume that $P_1$ is a small correction to this solution that is proportional to $\nu'_0$. 

Inserting this linear approximation into Eq.~(\ref{fpeus}) and equating terms of the same power of $\nu'_0$, one obtains for $0<\sigma<1$
\begin{align}
D \frac{\partial^2 P_0}{\partial \sigma^2} - \nu_0 P_0 + \Gamma\frac{e^{\sigma-1}}{\mathcal{N}} &= 0 , \nonumber \\
D \frac{\partial^2 P_1}{\partial \sigma^2} - e^{-\frac{1}{2^a T}} P_1 -\nu'_0 \left( \sigma - \frac{1}{2}\right) P_0 &= 0.\label{statdiffeqhight}
\end{align}
The first of these differential equations can be solved to yield a solution for $P_0$, which is simply 
\begin{align}
	P_0(\sigma) = \frac{\Gamma e^{\sigma -1}}{\mathcal{N}(\nu_0 - D )} + A_1 e^{\sqrt{\frac{\nu_0}{D}}\sigma} + A_2 e^{-\sqrt{\frac{\nu_0}{D}}\sigma} ,
\end{align}
where $A_1$ and $A_2$ are arbitrary constants of integration. This can then be inserted into the second of Eqs.~(\ref{statdiffeqhight}) to yield $P_1$. Putting these two solutions together, we obtain
\begin{align}
	P(\sigma) &\approx \frac{\Gamma e^{\sigma -1}}{\mathcal{N}(\nu_0 - D )} \left[ 1 +  \nu'_0 \frac{\nu_0 (1/2- \sigma) + D (\sigma-5/2)}{2(\nu_0 - D )^2}\right] \nonumber \\
	&+ A_1 e^{\sqrt{\frac{\nu_0}{D}} \sigma}\left[ 1 + \nu'_0 \frac{\sqrt{\nu_0 D} (1 - 2 \sigma) +2 \nu_0 \sigma (\sigma-1) + D }{8 \nu_0^{3/2} \sqrt{D} }\right]\nonumber \\
	&+ A_2 e^{-\sqrt{\frac{\nu_0}{D}} \sigma}\left[ 1 + \nu'_0 \frac{\sqrt{\nu_0 D} (1 - 2 \sigma) - 2 \nu_0 \sigma (\sigma-1) - D }{8 \nu_0^{3/2} \sqrt{D} }\right]
\end{align}
For $\sigma>1$ we instead have 
\begin{align}
	D \frac{\partial^2 P}{\partial \sigma^2} -  P  &= 0 ,
\end{align}
which yields simply
\begin{align}
	P(\sigma) = C e^{- \frac{\sigma}{\sqrt{D}}}. \label{outsidehight}
\end{align}
To find the constants $A_1$, $A_2$ and $C$, we require $\partial_\sigma P\vert_{\sigma = 0} = 0$ [which is necessary for $P(\sigma) = P(-\sigma)$], and continuity of $P(\sigma)$ and it's derivative at $\sigma = 1$. 

We note that we have not yet used the relationship between $D(\mathbf{r})$ and $\Gamma(\mathbf{r})$ to find the above solution for the stationary state $P(\sigma)$. By imposing that the solution for $P(\sigma)$ must be normalised, we find a constraint $Q(D, \Gamma)= 0$ that relates $D(\mathbf{r})$ and $\Gamma(\mathbf{r})$. By further imposing the mean-field approximation $D = \alpha \Gamma$, we obtain a unique homogeneous solution for $\Gamma$.

Expanding this solution for $\Gamma$ for large $T$, the full expression for which is long and uninformative, we find [as in Eq.~(\ref{gammahight}) of the main text]
\begin{align}
	\Gamma(T) &\approx 1 - \frac{T_\mathrm{onset}}{T}, \nonumber \\
	T_\mathrm{onset} &= \frac{\mathcal{N}}{(\alpha-1)(1-e)^2 2^{2+a}} e^{1 - \frac{2}{\sqrt{\alpha}}} \bigg[2 e^{\frac{2}{\sqrt{\alpha}}}(e - 1 - 3a + a e) - 2a \alpha^{3/2}\left(e^{\frac{1}{\sqrt{\alpha}}}-1\right)\left(2e^{\frac{1}{\sqrt{\alpha}}} + e^{1+\frac{1}{\sqrt{\alpha}}}-e\right) \nonumber \\
	& +(a-1)\sqrt{\alpha} e \left(e^{\frac{2}{\sqrt{\alpha}}} -1 \right) \nonumber \\
	&+ \alpha \left(e - 3 a e + 4a e^{1+ \frac{1}{\sqrt{\alpha}}} - (1 + a)e^{1+ \frac{2}{\sqrt{\alpha}}} + 2 (a-1) e^{\frac{1}{\sqrt{\alpha}}} + 2 (1+a) e^{ \frac{2}{\sqrt{\alpha}}}\right) \bigg].
\end{align}

\section{Solution for MCT-like regime} \label{appendix:mct}
We now approximate the function $\nu(\sigma, \sigma_c)$ as described in Eq.~(\ref{barrierapprox}). This approximation is compared with the true function in Fig. \ref{fig:nuapproxmct} below.

\begin{figure}[H]
	\centering 
	\includegraphics[scale = 0.5]{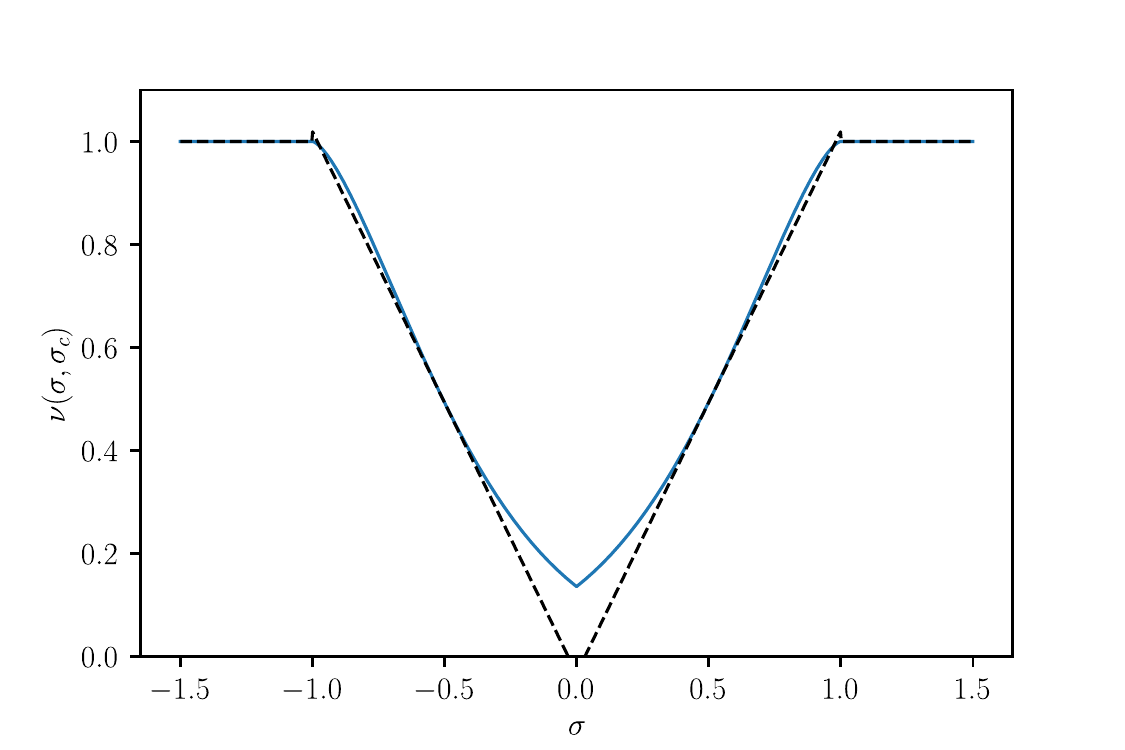}
	\captionsetup{justification=raggedright,singlelinecheck=false, font=small}
	\caption{Comparison of the approximate expression in Eq.~(\ref{barrierapprox}) with Eq.~(\ref{ratenu}) for $T = 0.5$.}\label{fig:nuapproxmct}
\end{figure}
In order to find $P(\sigma)$, we make two further approximations, the first of which is most useful for $T_c \lessapprox T$, and the other of which is better suited to higher temperatures.

\subsection{$T_c \lessapprox T$} 
In this case, we assume that the contribution to $P(\sigma)$ for $\sigma>1$ is negligible. For our piecewise solution for $P(\sigma)$ there are thus two regions of interest: $\vert \sigma \vert <\sigma_B(T)$ and $\sigma_B(T) < \vert\sigma\vert< 1$. 

The solution of Eq.~(\ref{fpeus}) for $0< \sigma <\sigma_B(T)$ is straightforward. One has
\begin{align}
	P(\sigma) = A_{11} + A_{12} \sigma -\Gamma \frac{e^{\sigma-1}}{D \mathcal{N}} .
\end{align}
For $\sigma_B(T) < \sigma< 1$, we instead write $P(\sigma) \approx P_0 + P_1 (\sigma-\bar\sigma) + P_2 (\sigma- \bar\sigma)^2$. This series solution is then substituted into Eq.~(\ref{fpeus}) and we obtain
\begin{align}
	P(\sigma) = A_{21} \left[ 1 + \frac{(\bar \sigma - \sigma_B)}{2 D} (\sigma - \bar \sigma)^2\right] + A_{22} (\sigma - \bar \sigma) + \Gamma\frac{e^{\bar\sigma-1}}{\Delta \mathcal{N}} \left[ 1 + \frac{1}{2} (\sigma - \bar \sigma) + \frac{\Delta}{2 D} (\bar \sigma - \sigma_B - 1) (\sigma- \bar \sigma)^2\right]. 
\end{align}
In this instance, the support of $P(\sigma)$ is confined to the region $\vert \sigma\vert < \sigma_{edge}$. In a similar manner as in Appendix \ref{appendix:highT}, one can find the constants $A_{11}$, $A_{12}$, $A_{21}$ and $A_{22}$ in terms of $\sigma_{edge}$ by requiring $\partial_\sigma P\vert_{\sigma = 0} = 0$, continuity of $P(\sigma)$ and $\partial_\sigma P$ at $\sigma = \sigma_B$ and that $P(\sigma_{edge}) = 0$. 

By subsequently imposing that $P(\sigma)$ be normalised \textit{and} imposing $\int_0^{\sigma_{edge}} \Delta(\sigma- \sigma_B) P(\sigma) = \Gamma$, one arrives at a constraint $Q(D, \Gamma) = 0$ relating $\Gamma$ and $D$. If one then expands this rather cumbersome but ultimately simple expression for small $\sigma_B-\bar \sigma$, small $D$ and small $\Gamma$, one arrives at the expression in Eq.~(\ref{gammaexpansion}), where the coefficients are given by
\begin{align}
	\beta_1 &= -\frac{4}{7}\frac{(e^{\bar\sigma}-1) \bar\sigma }{ e \mathcal{N}},\nonumber \\
	\beta_2(T) &= \frac{36}{49}\frac{ \bar \sigma}{ \Delta(T)}+ \frac{8}{21}\frac{(e^{\bar\sigma}-1)\bar\sigma}{e\mathcal{N}}.
\end{align}

\subsection{Series expansion around $\sigma_{1/2}$}
We use a similar approach here as in the previous subsection, but now we allow for the possibility that the contribution to $P(\sigma)$ for $\sigma>1$ is non-negligible, as it would be for $T$ appreciably higher than $T_c$.

Again, for $0< \sigma <\sigma_B(T)$ one has
\begin{align}
	P(\sigma) = A_{11} + A_{12} \sigma -\Gamma \frac{e^{\sigma-1}}{D \mathcal{N}} .
\end{align}
Now, for $\sigma_B<\sigma<1$, we instead seek a series solution for $P(\sigma)$ expanded about $\sigma_{1/2}$. One finds up to third order
\begin{align}
	P(\sigma) &= A_{21} \left[ 1 + \frac{1}{4 D} (\sigma - \sigma_{1/2})^2 + \frac{\Delta}{6 D} (\sigma-\sigma_{1/2})^3 \right] + A_{22}\left[ (\sigma - \sigma_{1/2}) + \frac{1}{12 D}(\sigma-\sigma_{1/2})^3\right] \nonumber \\
	& + \frac{1}{32}\frac{ \Gamma e^{\sigma_{1/2}-1}}{\mathcal{N}} \bigg[48 D \Delta^2 + 16 D^2 \Delta + 8 D \Delta - 2 D -1 + (16 D \Delta + 2\Delta - 2 D - 1)(\sigma-\sigma_{1/2}) \nonumber \\
	&- \frac{1}{2}(1 - 4 \Delta - 8 D \Delta + 8 \Delta^2)(\sigma-\sigma_{1/2})^2+ \frac{1}{6}(48 \Delta^3 + 16 D \Delta^2 - 24 \Delta^2 + 6 \Delta - 1)(\sigma-\sigma_{1/2})^3 \bigg].
\end{align}
For $\sigma>1$, we have once again
\begin{align}
	P(\sigma) = Ce^{-\frac{\sigma}{\sqrt{D}}}. 
\end{align}
In contrast to the previous subsection, the support of $P(\sigma)$ is now the entire real axis under this approximation scheme. One can solve for the constants of integration $A_{11}$, $A_{12}$, $A_{21}$, $A_{22}$ and $C$ in the usual manner. Upon imposing normalisation, one thus obtains the constraint $Q(D, \Gamma) = 0$. One imposes the mean-field assumption $D = \alpha \Gamma$ to find $\Gamma$ and thus obtain the theory lines in Fig. \ref{fig:statdistmc}.

\section{Solution for $T \to 0$}\label{appendix:smallt}
In this appendix, we derive the results presented in Section \ref{section:T0} of the main text, where we find the behaviour of the activity $\Gamma(\mathbf{r})$ for $T\to 0$, deriving both the mean-field value of $\Gamma$ and the typical lengthscale of inhomogeneities as a function of $T$. 

To this end, we introduced the approximation for $\nu(\sigma, \sigma_c)$ presented in Eq.~(\ref{nusmallt}). This approximation is compared with the true expression for $\nu(\sigma, \sigma_c)$ in Fig. \ref{fig:nuapproxlowt} below. We see that around $\sigma = \bar \sigma$, the approximation is a good fit. Closer to $\sigma =1$, the approximation breaks down. Since we expect the quantity $P(\sigma)$ to decay rapidly outside the region $\vert \sigma \vert <\bar \sigma$ for small $T$, we anticipate that the inaccuracy of $\nu(\sigma, \sigma_c)$ close to $\sigma = 1$ will not be of consequence.

\begin{figure}[H]
	\centering 
	\includegraphics[scale = 0.5]{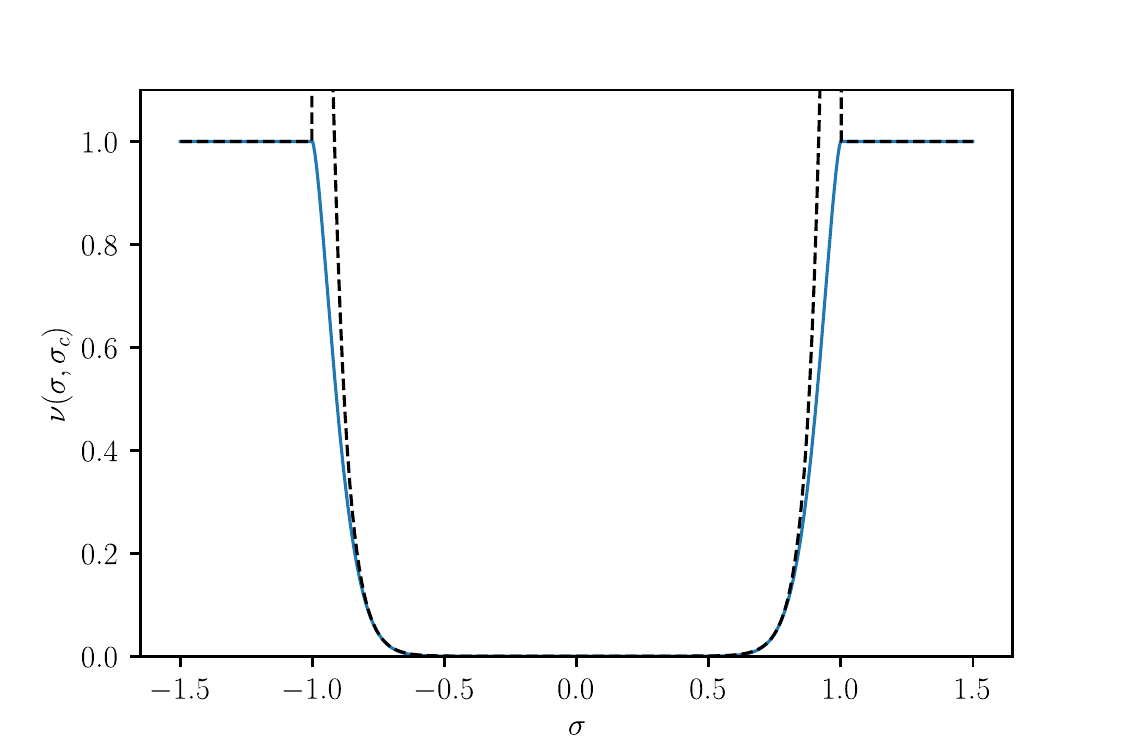}
	\captionsetup{justification=raggedright,singlelinecheck=false, font=small}
	\caption{Comparison of the approximate expression in Eq.~(\ref{nusmallt}) with Eq.~(\ref{ratenu}) for $T = 0.04$. }\label{fig:nuapproxlowt}
\end{figure}

We now provide the details on how one calculates the stationary distribution of stresses presented in Fig \ref{fig:statdist}a. This is done in a piecewise fashion. We first provide an approximate solution to Eq.~(\ref{fpeus}) for $0<\sigma <\bar \sigma$, and then discuss the solution for $\sigma>\bar\sigma$.
 
\subsection{Stationary solution for $P(\sigma)$}
For $0<\sigma < \bar\sigma$, we see that $e^{-\bar E/T -(\bar\sigma - \sigma)/T_1}/\Gamma\ll 1$. We thus suppose that in the region $0<\sigma < \bar\sigma$ we can decompose $P(\sigma) \approx P_0(\sigma) + P_1(\sigma)$, where $P_1(\sigma)$ is a correction of order $\sim e^{-\bar E/T}$. Equating terms of the same order of magnitude in Eq.~(\ref{fpeus}) with the expression for $\nu(\sigma,\sigma_c)$ given in Eq.~(\ref{nusmallt}), one finds
\begin{align}
	\frac{D}{\Gamma} \frac{\partial^2 P_0}{\partial \sigma^2} + \frac{1}{\mathcal{N}} e^{\sigma  - 1} &=0 , \nonumber \\
	\frac{D}{\Gamma} \frac{\partial^2 P_1}{\partial \sigma^2}  - \frac{e^{-\bar E /T} e^{-\frac{(\bar\sigma - \sigma)}{T_1}}}{\Gamma} P_0  &=0 .
\end{align}
Solving these in succession thus yields
\begin{align}
	P(\sigma) =& -\frac{\Gamma}{2 D^2 (e-1)}e^\sigma\left(D + \frac{T_1^2}{(1+T_1)^2} e^{-\bar E/T -(\bar\sigma - \sigma)/T_1}\right) + A_1 \left( 1 + \frac{T_1^2}{D} e^{-\bar E/T -(\bar\sigma - \sigma)/T_1}\right) \nonumber \\
	&\,\,\,\,\,+A_2 \left(\sigma + \frac{T_1^2}{D} (\sigma - 2 T_1)e^{-\bar E/T -(\bar\sigma - \sigma)/T_1}\right) ,\label{insidesolution}
\end{align}
where we note that the functions multiplying the arbitrary constants (i.e. the complementary functions) solve the homogeneous differential equation. 

Now, we consider the case $\sigma>\bar\sigma$, where we have $e^{-\bar E/T -(\bar\sigma - \sigma)/T_1}/\Gamma\gg 1$. In this instance, we must be careful about dealing with the complementary functions and the particular integral separately. The complementary functions satisfy
\begin{align}
	\frac{D}{\Gamma} \frac{\partial^2 P_1}{\partial \sigma^2}  - \frac{e^{-\bar E /T} e^{-\frac{(\bar\sigma - \sigma)}{T_1}}}{\Gamma} P_1 &= 0. \label{cfdiff}
\end{align}
The full solutions to this equation are modified Bessel functions of the Arrhenius factor. However, since we are only interested in the low-$T$ behaviour, we can approximate the complementary function as
\begin{align}
	P_{CF} = B_1 e^{\frac{(\bar\sigma - \sigma)}{2 T_1}} \exp\left[-\frac{2 T_1 e^{-\bar E /(2T)} e^{-\frac{(\bar\sigma - \sigma)}{2T_1}} }{\sqrt{D}} \right] ,
\end{align}
which one can see satisfies Eq.~(\ref{cfdiff}) up to leading order in the Arrhenius factor. We note that we have ignored the other possible solution of Eq.~(\ref{cfdiff}), which would blow up very quickly for increasing $\sigma$. 
	
When considering the particular integral of Eq.~(\ref{fpeus}) in the regime $\sigma > \bar\sigma$, we instead ignore the second derivative, which is now small in comparison to the other terms. For small $T$, we find simply 
\begin{align}
	P_{PI} = \frac{\Gamma }{\mathcal{N}}e^{\bar E /T} e^{\frac{(\bar\sigma - \sigma)}{T_1}} e^{ \sigma  - 1},
	\end{align}
as was similarly suggested in Ref. \cite{ozawa2023elasticity}. So the full approximate solution for $\sigma>\bar\sigma$ is 
\begin{align}
	P(\sigma) =  \frac{\Gamma }{\mathcal{N}}e^{\bar E /T} e^{\frac{(\bar\sigma - \sigma)}{T_1}} e^{ \sigma  - 1} + B_1 e^{\frac{(\bar\sigma - \sigma)}{2 T_1}} \exp\left[-\frac{2 T_1 e^{-\bar E /(2T)} e^{-\frac{(\bar\sigma - \sigma)}{2T_1}} }{\sqrt{D}} \right].\label{outsidesolution}
\end{align}
Once again, we must find the constants of integration $A_1$, $A_2$ and $B_1$. This is accomplished by imposing that 1) the derivative of $P(\sigma)$ at $\sigma = 0$ must be zero, 2) the solution must be continuous at $\sigma = \bar\sigma$, 3) the solution must have a continuous derivative at $\sigma = \bar\sigma$. One thus arrives at a solution of Eq.~(\ref{fpeus}) that ought to be valid for small $T$ for a fixed combination of $D$ and $\Gamma$.
	
	
	
\subsection{Scaling of the mean field activity $\Gamma(T)$}\label{appendix:gammaTdependenceT0}
By requiring that the solution for $P(\sigma)$ be normalised, we again obtain a constraint $Q(\Gamma, D) = 0$. As mentioned in the main text, if we make the substitutions $d = D e^{\bar E/T}$ and $\gamma = \Gamma e^{\bar E/T}$, we obtain a substantially simpler expression for the constraint $q(\gamma, d) = 0$, where $q(\cdot,\cdot)$ is a rational function of its arguments. Specifically, we have
\begin{align}
	\gamma & \Bigg\{ (1-T_1)(1+T_1)^3 \bigg[2 d^2 (s^2 -2)T_1 + 10 d^{1/2} T_1^6 + 4 T_1^7 + d^{5/2} (s^2 + 4 s T_1 -2) \nonumber \\
	&+ 2 d T_1^{3} (s^2 - 4 s T_1 + 6 T_1^2 -2) + d^{3/2} T_1^2 (3s^2 - 8 s T_1 + 10 T_1^2 -6)\bigg]\nonumber \\
	&+2 e^{\bar \sigma}\bigg[(t_1 -1) t_1^7 + 2 d^3 (T_1 s - s - T_1)T_1 (1+T_1)^3 + d^{1/2} T_1^6 (3 T_1 2 T_1^2-5) + d^{7/2}(1+T_1)^3(s - T_1 - 3 sT_1+2 sT_1^2)\nonumber \\
	&+ 2 d^2 T_1 (1+T_1)^3 (1 + sT_1 -s - T_1 -2 T_1^3 +T_1^4)+ 2d (T_1-1)T_1^3(s-3T_1 +s T_1 +2 T_1^3 + T_1^4-1) \nonumber \\
	&+d^{5/2} (1 + T_1)^3(1 - T_1 - 2 T_1^3 - 3 T_1^4 +2 T_1^5 -s -sT_1 +2s T_1^2)\nonumber \\
	&+d^{3/2}(T_1 -1)T_1^2(6T_1^3 + 7T_1^4 +2 T_1^5 +3s +5sT_1+2sT_1^2-4T_1^2-9T_1-3)\bigg]\Bigg\}\nonumber \\
	&= 2(e -1)  (T_1 -1)(1+T_1)^3  d^2 \left(d^{3/2} + 2 d T_1 + 3 d^{1/2} T_1^2 + 2 T_1^3\right). \label{qconstraint}
\end{align}
Rearranging for $\gamma$ and expanding this expression for small $d$ and $T_1$, one finds the approximate expression given in Eq.~(\ref{constraintlowt}) in the main text with
\begin{align}
	c_1 &= - \frac{2(e^{\bar\sigma-1})\bar\sigma}{\alpha^2 (e-1)},\nonumber \\
	c_2 &= - \frac{4 e^{\bar\sigma}\bar\sigma}{\alpha^2 (e-1)}, \nonumber \\
	c_3 &= \frac{e^{\bar\sigma}\bar\sigma}{\alpha^2 (e-1) }.
\end{align}
We test the scaling behaviour of the mean-field value of $\gamma$ for $T\to 0$ [obtained by solving Eq.~(\ref{constraintlowt}) simultaneously with $d = \alpha \gamma$ -- see Eq.~(\ref{gammascaling})] in Fig. \ref{fig:scalingtest} below.

\begin{figure}[H]
	\centering 
	\includegraphics[scale = 0.5]{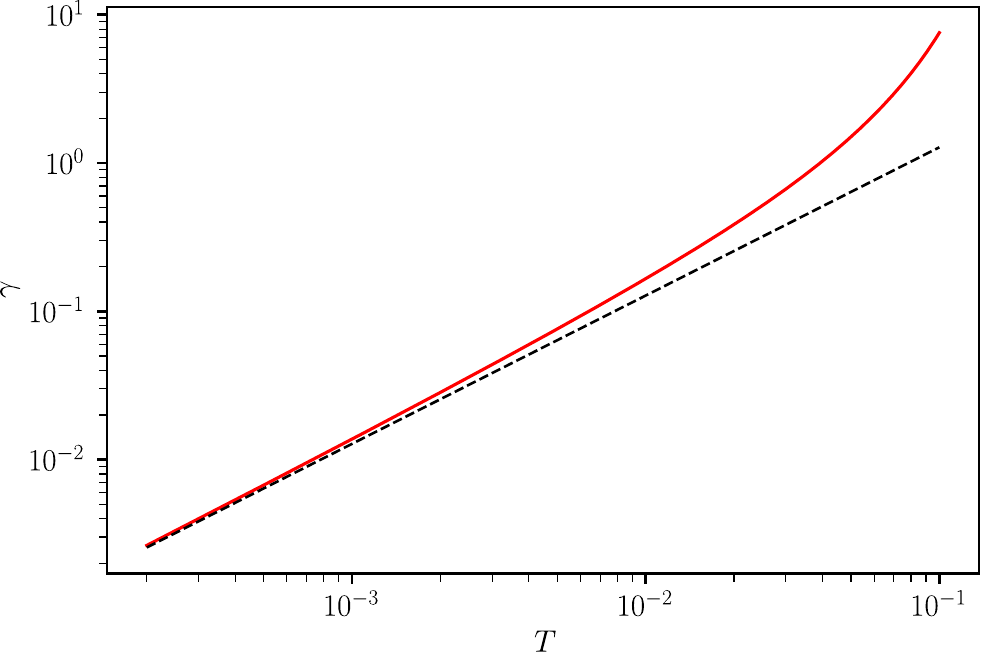}
	\captionsetup{justification=raggedright,singlelinecheck=false, font=small}
	\caption{The numerical solution of Eq.~(\ref{qconstraint}) with $d = \alpha \gamma$ (solid red line) is compared to the asymptotic behaviour given in Eq.~(\ref{prefactorscaling}) (dashed black line) of the main text. }\label{fig:scalingtest}
\end{figure}

\section{Simulation of the mean-field dynamics}\label{appendix:numerics}
We perform simulations of the mean-field dynamics using a variant of the Gillespie algorithm. We take an ensemble of $N$ identical sites, each of which have an associated stress $\sigma_i$, such that $i = 1, \cdots N$. Each stress may undergo plastic rearrangements at a rate $\nu(\sigma_i, \sigma_c)$. When a stress relaxes, and is subsequently redrawn from the distribution $y(\sigma_i)$, each of the other receives a stress `kick' $\delta \sigma_i$. Each kick is drawn independently from a Gaussian distribution such that $\delta \sigma_i \sim \mathcal{N}\left(0, \frac{2\alpha}{N}\right)$. For large $N$, the distribution of stresses $P_i(\sigma_i, t)$ ought to be that described by Eq.~(\ref{fpegeneral}).
	
More precisely, the algorithm can be summarised as follows
\begin{itemize}
	\item Initialise the system by drawing each stress from some starting distribution $P_0(\sigma_i)$. We choose a Gaussian distribution $P_0(\sigma) = \frac{1}{\sqrt{2\pi\sigma_c^2}}\exp\left(-\frac{\sigma^2}{2\sigma_c^2} \right)$. For non-zero $T$, the long-term behaviour is independent of this choice. Set the time $t = 0$.
	\item Calculate a rate of plastic events for each stress $r_i = \nu(\sigma_i, \sigma_c)$. Calculate the total rate $R = \sum_i r_i$.
	\item Obtain the time until the next event ($\delta t$) by drawing a uniform random number ($u_1$) that can take values between 0 and 1. The waiting time is given by $\delta t = -R^{-1} \ln(u_1)$. Update $t \to t + \delta t$.
	\item Choose which stress undergoes the plastic event. Draw another uniform random number between 0 and 1, $u_2$. Defining $c_i = \sum_{j = 1}^i\frac{r_j}{R}$, find $i$ such that $c_{i-1}< u_2 \leq c_i$ (where $c_0 = 0$). Redraw $\sigma_i$ from the distribution $y(\sigma)$.
	\item Give every other stress a kick. Update each stress by adding a Gaussian random number such that $\sigma_i \to \sigma_i + \delta \sigma_i$ and each kick is drawn independently such that $\delta \sigma_i \sim \mathcal{N}\left(0, \frac{2\alpha}{N}\right)$. 
	\item Repeat from the second step.
\end{itemize}

\end{widetext}

\end{document}